\documentclass[preprint,aps]{revtex4}
\usepackage[latin9]{luainputenc}
\usepackage{float}
\usepackage{amsmath,amssymb}
\usepackage{color}
\usepackage{amssymb}
\usepackage{graphicx}
\usepackage{setspace}
\usepackage{hyperref}
\makeatletter

\@ifundefined{textcolor}{}
{%
 \definecolor{BLACK}{gray}{0}
 \definecolor{WHITE}{gray}{1}
 \definecolor{RED}{rgb}{1,0,0}
 \definecolor{GREEN}{rgb}{0,1,0}
 \definecolor{BLUE}{rgb}{0,0,1}
 \definecolor{CYAN}{cmyk}{1,0,0,0}
 \definecolor{MAGENTA}{cmyk}{0,1,0,0}
 \definecolor{YELLOW}{cmyk}{0,0,1,0}
}

\usepackage{amsfonts}
\usepackage{bm}
\usepackage[dvips]{epsfig}
\usepackage[dvips]{graphicx}
\usepackage{float}
\usepackage{dcolumn}
\usepackage{ifpdf}
\usepackage{multirow}

\newcommand{\be}{\begin{equation}}
\newcommand{\ee}{\end{equation}}
\newcommand{\bes}{\begin{subequations}}
\newcommand{\ees}{\end{subequations}}
\newcommand{\ben}{\begin{eqnarray}}
\newcommand{\een}{\end{eqnarray}}

\usepackage{hyperref}
\hypersetup{
    colorlinks=true, 
    linkcolor=black,
    citecolor=black
}
\DeclareMathOperator{\sign}{sign}

\usepackage{soul}

\makeatother

\begin{document}

\title{Half-line kink scattering in the $\phi^4$ model with Dirichlet boundary conditions}
\date{}

\author{Jairo S. Santos$^1$, Fabiano C. Simas$^{1,2}$, Adalto R. Gomes$^{1,2}$}
\email{jairo.ss@discente.ufma.br, fc.simas@ufma.br, adalto.gomes@ufma.br}

\affiliation{$^1$ Departamento de F\'{\i}sica, Universidade Federal do Maranh\~ao Campus Universit\'ario do Bacanga, 65085-580, S\~ao Lu\'{\i}s, Maranh\~ao, Brazil}

\affiliation{$^2$ Programa de P\'os-Gradua\c c\~ao em F\'\i sica, Universidade Federal do Maranh\~ao\\Campus Universit\'ario do Bacanga, 65085-580, S\~ao Lu\'\i s, Maranh\~ao, Brazil.
}
%\author{}

%\selectlanguage{english}

%%%%%%%%%%%%%%%%%%%%%%%%%%%%%%%%%%%%%%%%%%%%%%%%%%%%%%%%%%%%%%%%%%%%%%%%%%%%%
\begin{abstract}
In this work, we investigate the dynamics of a scalar field in the nonintegrable $\displaystyle \phi ^{4}$ model, restricted to the half-line. Here we consider singular solutions that interpolate the Dirichlet boundary condition $\phi(x=0,t)=H$ and their scattering with the regular kink solution. The simulations reveal a rich variety of phenomena in the field dynamics, such as the formation of a kink-antikink pair,  the generation of oscillons by the boundary perturbations, and the interaction between these objects and the boundary, which causes the emergence of boundary-induced resonant scatterings (for example, oscillon-boundary bound states and kink generation by oscillon-boundary collision) founded into complex fractal structures.
Linear perturbation analysis was applied to interpret some aspects of the scattering process.  We detected the presence of two shape modes near the boundary. The power spectral density of the scalar field at a fixed point leads to several frequency peaks. Most of them can be explained with some interesting insights for the interaction between the scattering products and the boundary. 

\end{abstract}

\maketitle

%%%%%%%%%%%%%%%%%%%%%%%%%%%%%%%%%%%%%%%%%%%%%%%%%%%%%%%%%%%%%%%%%%%%%%%%%%%%%
\section{introduction}
%%%%%%%%%%%%%%%%%%%%%%%%%%%%%%%%%%%%%%%%%%%%%%%%%%%%%%%%%%%%%%%%%%%%%%%%%%%%%

The kink (antikink) is the simplest topological defect. It is a solution of a $(1,1)$ scalar field theory that interpolates between two different energy minimum solutions. The topological aspect means that there is a mapping between the physical space and the field configuration space described by an integer named the topological charge \cite{vachas}. A kink has a topological charge $+1$, while an antikink has a topological charge $-1$. The energy density distribution of a static kink is localized. A traveling kink is characterized by free propagation without loosing form, that is, it is a solitary wave. 

The present work considers the $\phi^4$ model, a well-known and much-studied field theory. In relation to symmetry breaking, the $\phi^4$ model had several applications in physics. One can cite chiral transition in QCD \cite{qcd}, Higgs mechanism \cite{part1}, domain structure of the vacuum in early cosmology \cite{cosm1},  non-local quantum field theories \cite{nonl}, superconductivity \cite{cm1} and ferroelectrics \cite{cm2}. From the point of view of nonlinear science and solitary waves, the $\phi^4$ model has a long history of investigation and is still an object of investigation \cite{phi4}.    

Being a nonintegrable model, the kink-antikink scattering from $\phi^4$ model shows a complex pattern. This includes bion states for low initial velocities $v$ of the pair, one-bounce collision for high $v$, and two-bounce collisions for intermediate velocities. Moreover, there appears three-bounce collisions close to two-bounce collisions and so on for $n>2$, in a fractal pattern \cite{pat1}. For a more historical account of the early investigations on these subject, see the Ref. \cite{camp}. The resonance structure of kink-antikink scattering and a phenomenological explication for bounce windows was presented in the Ref. \cite{k1}. The use of collective coordinates as an effective model to reproduce the main aspects of scattering was presented in the Ref. \cite{suj}.  However, collective coordinates as a superposition of the kink, antikink, and mode profiles leads to a singularity of the collective coordinate approximation when kink and antikink get arbitrarily close. This is called the null-vector problem \cite{cap}, which demands more suitable parametrizations \cite{k4,k5}. For the $\phi^4$ model, this problem has been solved in the Refs. \cite{m1,m2}. The analysis of the interaction of the BPS mode with the excited mode led to the discovery of the spectral wall, which is a spatially localized region, defined by the point where an oscillation mode enters the continuous spectrum \cite{adam}. 
Other interesting aspects of investigation in the $\phi^4$ model include the interaction between kink and radiation \cite{rad1} leading to negative pressure \cite{rad2}, and meson-kink scattering resulting in positive pressure \cite{mes}.

There are very few examples in the literature of the effect of nonintegrability in the half-line, such as studies with the sine-Gordon model with Robin boundary conditions (b.c.) \cite{sg} and the $\phi^6$ model with Neumann b.c. \cite{phi6}. An interesting issue of the $\phi^4$ model is the dynamics restricted to a half line by a simple Neumann-type ``magnetic field" b.c. \cite{dor}. Such a boundary is a natural deformation of full-line scattering problems with, for some parameters, similar resonant scattering results. Also, there appears interesting behaviors such as `sharp-edged' scattering window, the creation of kinks by an excited boundary and a sudden burst of radiation from the boundary as a new decay channel \cite{dor}. In the Ref. \cite{chered}, the theory of particle scattering on a half-line was discussed, leading to factorizing conditions. The Ref. \cite{mackay} examined integrable boundary conditions on the half-line for the chiral model. In Ref. \cite{mackay1}, the SO(N) chiral model on the half-line resulted in an infinite number of conserved charges by combining Dirichlet and Neumann boundary conditions. Finally, it is worth commenting on the Kondo model in a three-dimensional nonrelativistic problem. Restricting to radial coordinates, the massless fermions move in the half-line, with the impurity at the boundary \cite{fend}

In the present work, we will consider the $\phi^4$ model with Dirichlet boundary condition, showing that new interesting aspects do appear. In the next section we review the model, its regular and irregular solutions on the full line and static solutions that connect the boundary conditions with the positive minimum of the potential. The section \ref{sec3} presents the numerical results of kink-boundary scattering. In the Sect. \ref{sec4}, we explore the stability analysis and obtain the boundary modes.  In the Sect. \ref{sec5} we present some resonant scatterings. This includes the kink-boundary resonance mechanism, bounce scattering of oscillating pulses and oscillon-boundary interaction.  In the Sect. \ref{sec6} we present our main conclusions.  Supplementary material are attached and described in the Sect. \ref{sec7}.

%%%%%%%%%%%%%%%%%%%%%%%%%%%%%%%%%%%%%%%%%%%%%%%%%%%%%%%%%%%%%%%%%%%%%%%%%%%%%
\section{The model} \label{sec2}
%%%%%%%%%%%%%%%%%%%%%%%%%%%%%%%%%%%%%%%%%%%%%%%%%%%%%%%%%

%%%%%%%%%%%%%%%%%%%%

We consider a $\varphi^4$ theory, described by the action
\begin{equation}
    S = \int d^2x \left[\tfrac{1}{2}\partial_\mu \varphi\partial^\mu \varphi - V(\varphi)\right],
\end{equation}
with the potential $V(\varphi) = \tfrac{1}{2}(\varphi^2 - 1)^2$, whose minima are $\varphi_v = \pm 1$. This is a nonintegrable theory that supports energetically stable and time-independent solutions. Minimizing the action, the equation of motion  is the non-linear wave equation,
\begin{equation}
    \partial _{t}^{2} \varphi -\partial _{x}^{2} \varphi =V_{\varphi }( \varphi ),
    \label{eq:motion}
\end{equation}
where $V_{\varphi}=dV/d\varphi$. The trivial solutions of the motion equation, $\varphi=+1$ and $\varphi=-1$ correspond to the minima of the potential.  Static and stable solutions that interpolate between adjacent minima are also solutions of the first-order equation 
\begin{equation}
    \frac{d\varphi }{dx} =\pm \sqrt{2V( \varphi )}.
    \label{eq:bogomolnyi}
\end{equation}
This leads to the kink $\varphi_{K}(x) = \tanh (x)$ and antikink $\varphi_{\Bar{K}}(x) = -\varphi_{K}(x) = -\tanh(x)$ solutions.  Other solutions $\varphi(x) = \pm\coth(x)$, are commonly ignored in the analysis of the scalar field dynamics on the full line, since their total energy diverge due to their behavior in the limit $x\to 0$. 

 %%%%%%%%%%%%%%%%%%%%%%%%%%%%%%%%%%%%%%%%%%%%%%%%%%%%%%%%%%%%%%%%%%%%%%%%%%%%%%%%%%%%%%%%%%%%%%%%%%%%%%%%%%%%
\begin{figure}[t]
    \centering
    \includegraphics[width=\textwidth]{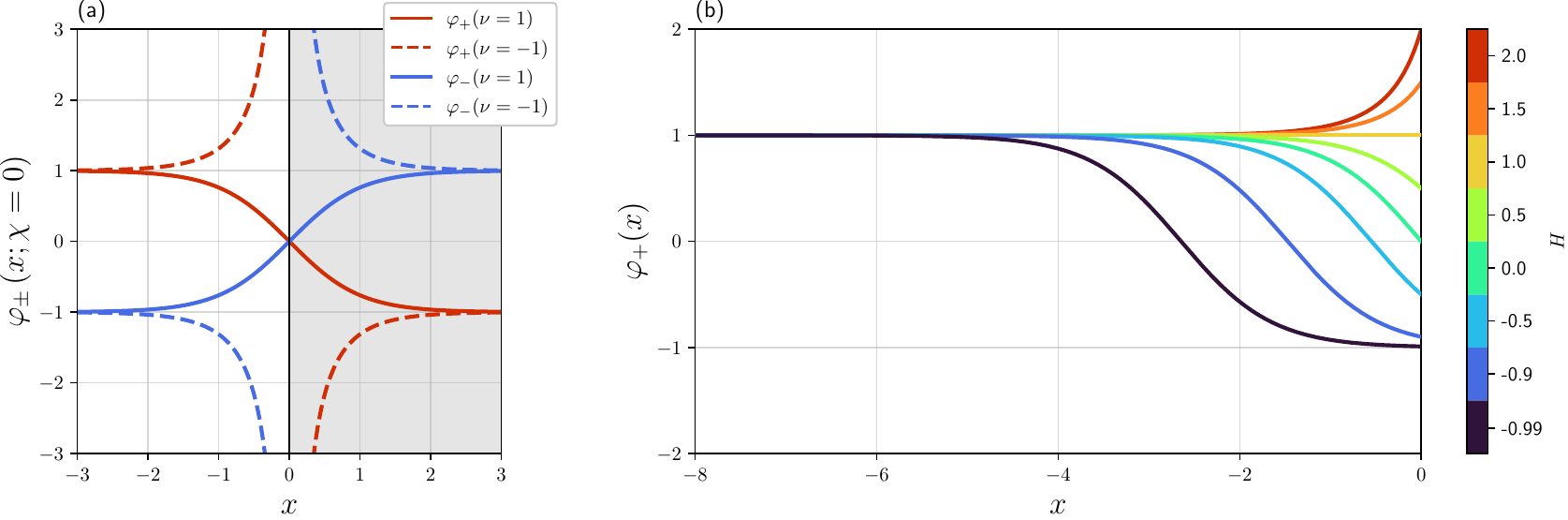}
    \caption{(a) All the solutions of the Bogomolnyi equation for $\chi = 0$ and (b) some solutions that connect the boundary condition with the positive vacuum.}
    \label{fig:solutions}
\end{figure}
%%%%%%%%%%%%%%%%%%%%%%%%%%%%%%%%%%%%%%%%%%%%%%%%%%%%%%%%%%%%%%%%%%%%%%%%%%%%%%%%%%%%%%%%%%%%%%%%%%%%%%%%%%%%

In this work, we consider the system restricted to the half-line $x\in(-\infty,0]$ with a Dirichlet boundary condition $\varphi(x,t) = H$ at $x=0$, so we consider regular kinks to the half-line and irregular kink on the full line. For this, let use a convenient form of the Bogomolnyi equation solutions given by 
\begin{equation}
    \varphi_{\pm} (x) =\pm(\tanh(\pm\chi - x))^{\nu },
    \label{eq:solutions}
\end{equation}
where $\nu\in\left\{\pm 1\right\}$ and $\chi=\chi(H)$. 

As we can see in the Fig. \ref{fig:solutions}a, correctly adjusted spatial displacements $\chi$ in $\varphi_{+}$ can access $H > -1$ without causing divergence, in the same way that $\varphi_{-}$ can intercept $H < 1$. 
Therefore, it is impossible to build a finite-energy initial condition for kink-boundary scattering when the Dirichlet condition imposes $H \leq -1$. Symmetrically, do not exist finite-energy solutions that allow antikink-boundary scattering if $H \geq 1$. Note as well that the $\nu$ parameter has to be chosen accordingly with the boundary value $H$, and this relation is given by $\nu = \sign(1 - |H|)$. Applying the boundary condition, we found 
\begin{equation}
 \chi = \tanh^{-1} \left(H^{\nu }\right)   
 \label{eq:displacements}
\end{equation}
as displacements that provide the Dirichlet condition for the Eq. (\ref{eq:solutions}). Thus, the positive solution $\varphi_{+}$ give us the stable connection between boundary and the positive vacuum, as the same way that negative solution $\varphi_{-}$ bind boundary and the negative vacuum. In the Fig. \ref{fig:solutions}b, we can see the solutions that interpolate between $\varphi=1$ and the boundary value $\varphi=H$ with the displacement $\chi$ given by the  Eq. (\ref{eq:displacements}). As we can clearly see, as one approaches $H=-1$ the positive boundary solution get an antikink form. Symmetrically, the negative boundary solution $\varphi_-$ becomes a kink as $H \to 1$. In general, $\varphi_\pm(x;H\to\mp1) \to \varphi_{\Bar{K},K}(x\mp\chi)$.
%%%%%%%%%%%%%%%%%%%%%%%%%%%%%%%%%%%%%%%%%%%%%%%%%%%%%%%%%%%%%%%%%%%%%%%%%%%%%%%%%%%%%%%%%%%%%%%%%%%%%%%%%%%%
\section{Numerical results} \label{sec3}
%%%%%%%%%%%%%%%%%%%%%%%%%%%%%%%%%%%%%%%%%%%%%%%%%%%%%%%%%%%%%%%%%%%%%%%%%%%%%%%%%%%%%%%%%%%%%%%%%%%%%%%%%%%%
In this section we describe the results of kink-boundary scattering. The numerical simulations were performed to solve the Eq. (\ref{eq:motion}) over a discrete spatial domain, limited at the range $[-L,0]$, where $L = 200$, with $N = 2048$ nodes equally spaced by $\delta x = L/(N-1) \approx 0.1$. We assume that the kink starts at $x_0=-10$ and moves toward the boundary. Aiming the differentiation, the 4\textsuperscript{th} order finite-difference method was applied with ghost-cells at the edges for reflective condition, at $x=-L$, and the Dirichlet condition at $x=0$. Moreover, we use 4\textsuperscript{th} order Runge-Kutta integrator over the temporal dimension.

The initial field configuration was been defined for analyze the scattering of a moving kink by the Dirichlet boundary, thus we took
\begin{equation}
    \left\{
    \begin{array}{l}
        \phi(x,0) = \varphi_K(\gamma(x - x_0)) + \varphi_{+}(x;H) - 1\\
        \dot{\phi}(x,0) = -\gamma v\left.\frac{d\varphi_K(\xi)}{d\xi}\right|_{\xi=\gamma(x-x_0)}%\varphi'_K(\gamma(x - x_0))
    \end{array}
    \right.
\end{equation}
for $v \in [0,1)$, $H > -1$ and $\gamma=(1-v^2)^{-1/2}$. Similarly, the negative boundary solution $\varphi_{-}$ can be used to define initial antikink-boundary configuration, but there is no loss of generality since the potential has only two minima and $V(\varphi)=V(-\varphi)$.
%%%%%%%%%%%%%%%%%%%%%%%%%%%%%%%%%%%%%%%%%%%%%%%%%%%%%%%%%%%%%%%%%%%%%%%%%%%%%%%%%%%%%%%%%%%%%%%%%%%%%%%%%%%%
\begin{figure}[t]
    \centering
    \includegraphics[width=\textwidth]{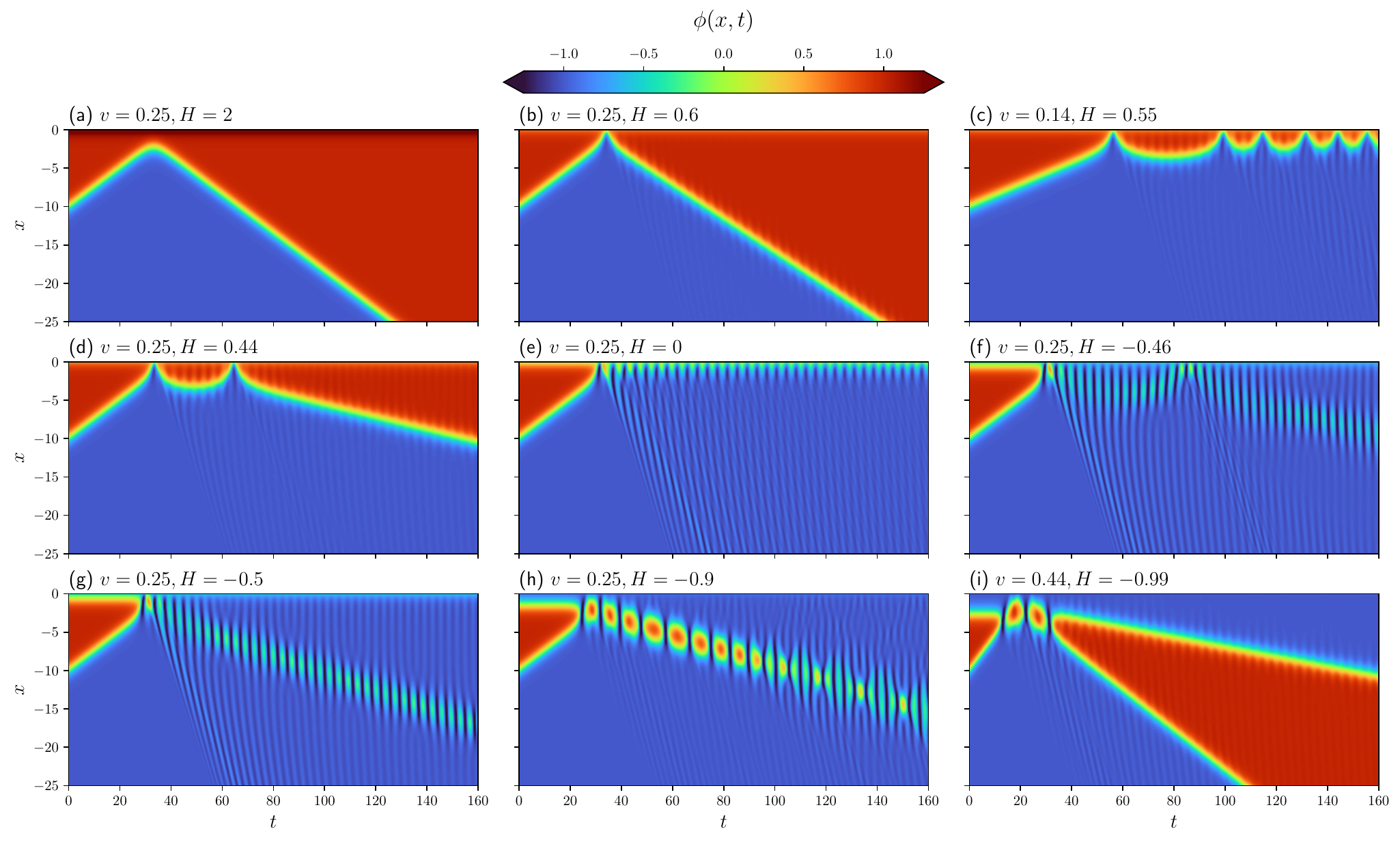}
    \caption{Kink-boundary scattering. (a) elastic scattering; (b) one-bounce phenomenon; (c) a bound-state between the kink and boundary; (d) two-bounce phenomenon; (e) total annihilation of the kink, in which case most of the energy decays into boundary perturbations; (f) an oscillon is released, returning and colliding once before moving away from the boundary; (g) emission of an oscillon; (h) production of a moving bion; (i) quasi three-bounce phenomenon in the interaction between the incoming kink and the released antikink. The videos depicting the scattering of figures (a), (d) and (e) are included in the \hyperref[sec7]{Supplementary Material} as video1.mp4, video2.mp4 and video3.mp4, respectively.}
    \label{fig:summary}
\end{figure}
%%%%%%%%%%%%%%%%%%%%%%%%%%%%%%%%%%%%%%%%%%%%%%%%%%%%%%%%%%%%%%%%%%%%%%%%%%%%%%%%%%%%%%%%%%%%%%%%%%%%%%%%%%%%

The Fig. \ref{fig:summary} depicts the main representative results $x(t)$ of the kink-boundary scattering for some values of $v$ and $H$. For $H \geq 1$ no significant level of radiation was observed, and we have almost elastic scattering (see the Fig. \ref{fig:summary}a). The kink-boundary interaction becomes attractive as we decrease the boundary parameter to $H \lesssim 1$ (see the Fig. \ref{fig:summary}b-i). For $H<1$ we have more evidence of inelastic scattering, with radiation released after the collision. 

If the kinetic energy of the kink is sufficiently high , it will produce a one-bounce situation, as shown in the Fig. \ref{fig:summary}b. Kinks with velocities lower than a critical velocity $v_c$ may not escape from the boundary, due to the attractive force, becoming a bound state that rapidly decays into only boundary oscillations - see the Fig. \ref{fig:summary}c. In the Fig. \ref{fig:summary}d the field bounces twice at the boundary before the kink escapes from the attraction, emitting some radiation. For velocities $v \lesssim v_c$, we found a fractal pattern of resonance windows in the kink-boundary interaction, analog to the full-line kink-antikink scattering. Decreasing the velocity at $v < v_c$, the output is the total annihilation of the kink, as illustrated in Fig. \ref{fig:summary}e, followed by oscillations close to the boundary. In the Fig. \ref{fig:summary}f, an oscillatory pulse is generated when the kink is close to the boundary. The oscillations then separate from the boundary but do not have enough energy to escape from a new collision. When we vary the value of the $H$ for fixed $v$, the oscillations are emitted with almost constant velocity - see the Fig. \ref{fig:summary}g.  In this case, we observe the emission of an oscillon. This structure corresponds to a low dispersive pulsating configuration around one minimum, whose small amplitude does not reach the local maximum of the potential during dynamics. In the Fig. \ref{fig:summary}h the emitted oscillations have a bion aspect,  where the field configuration is capable of oscillating between the minima of the model. Note that in this case, the pair oscillates without a recognizable pattern until their annihilation.  Finally, in the Fig. \ref{fig:summary}i we see that there is the emission of a kink-antikink from the boundary after  the kink have bounced three times at the boundary. 

%%%%%%%%%%%%%%%%%%%%%%%%%%%%%%%%%%%%%%%%%%%%%%%%%%%%%%%%%%%%%%%%%%%%%%%%%%%%%%%%%%%%%%%%%%%%%%%%%%%%%%%%%%%
\begin{figure}[t]
    \centering
    \includegraphics[width=\linewidth]{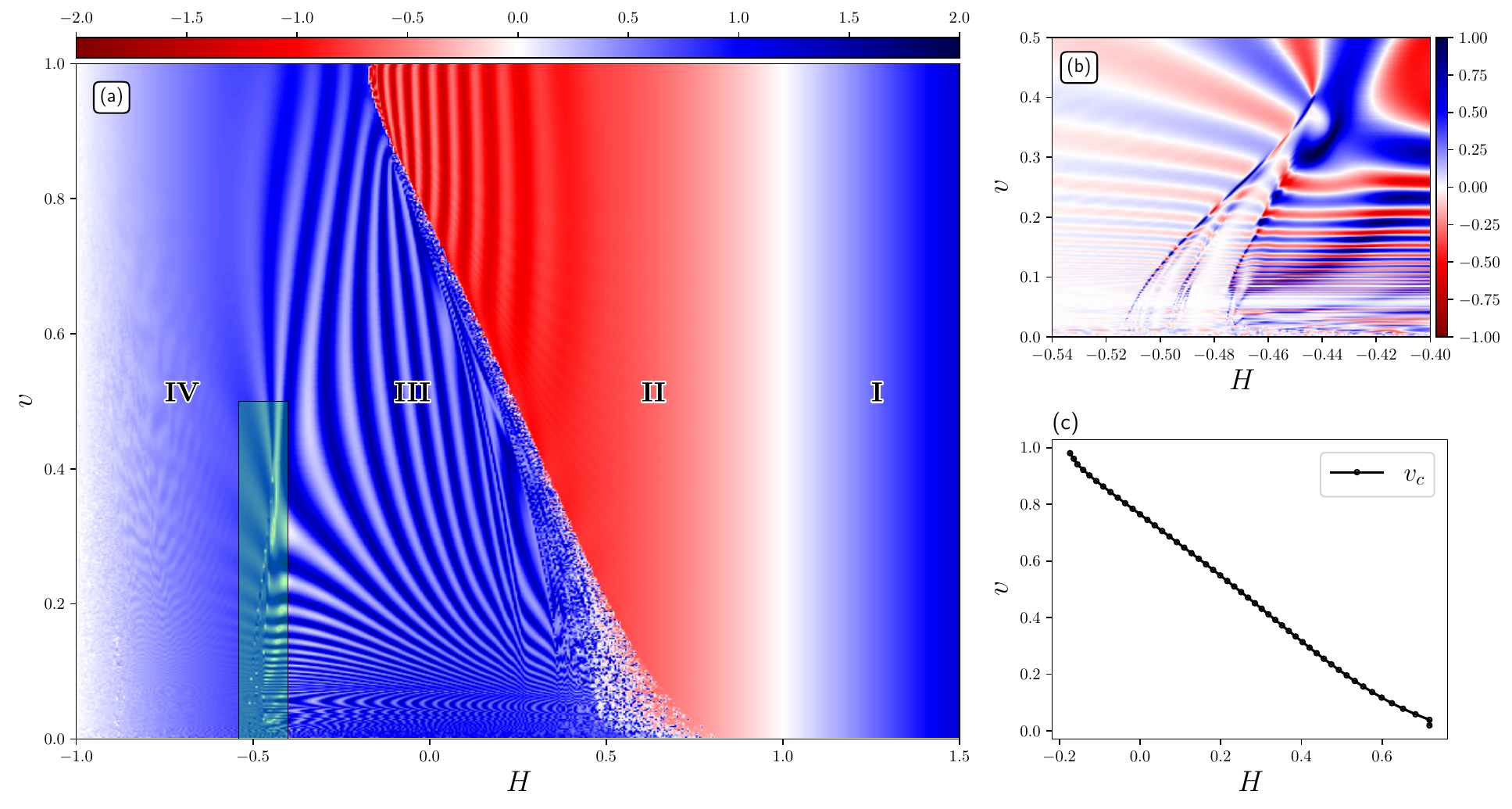}
    \caption{(a) The first spatial derivative $\phi'(x,t)$ measured at $x=0$ and $t=|x_0|/v + 100$,  showing four main scenarios. Region I: elastic scattering; region II: inelastic scattering (one-bounce); region III: boundary-induced annihilation; and region IV: oscillon/bion creation. (b) Measurements of $\phi'(x,t) - \varphi'_{-}(x)$ at $x=0$ into the green region showing  the oscillon-boundary resonance windows (the ``branches''-like pattern). (c) The critical velocity as a function of $H$.}
    \label{fig:mosaico}
\end{figure}
%%%%%%%%%%%%%%%%%%%%%%%%%%%%%%%%%%%%%%%%%%%%%%%%%%%%%%%%%%%%%%%%%%%%%%%%%%%%%%%%%%%%%%%%%%%%%%%%%%%%%%%%%%%%

The Fig. \ref{fig:mosaico}a shows the first derivative $\phi'(x,t)$ measured at the boundary quite a long time after the scattering in the phase space $v$ versus $H$. In the figure we characterize the main results of the scattering. In the region (I) we have an almost elastic scattering (see also the Fig. \ref{fig:summary}a). The region (II) represents one-bounce scattering  (Fig. \ref{fig:summary}b). Note also that the boundary between regions (II) and (III) characterizes the critical velocity, $v_c$, and the transition between these two regions occurs only for $-0.2 \lesssim H \lesssim 0.8$. We note that for $0 \leq H \lesssim 0.5$, we have a linear dependence of $v_c$ with $H$. This is more evident in the Fig. \ref{fig:mosaico}c. In the region (III), we have boundary-induced kink annihilation, which is characterized by a trapped oscillation state near the boundary (Fig. \ref{fig:summary}e), and for the eventual emission of oscillation pulses for high $v$. In the region (IV), we have mostly oscillation pulses and bion propagation (Figs. \ref{fig:summary}g-h). In the transition region between (III) and (IV), we have the presence of oscillating pulses that suffer bounce scattering with the boundary (Fig. \ref{fig:summary}f).  The Fig. \ref{fig:mosaico}b shows a zoomed-in of this region. In the figure, the white regions characterize the parameters where $\phi'$ is low, meaning that the boundary had loosen energy to the emitted oscillon. For $H \ll -0.5$, the oscillon size increases as the  lower is the boundary parameter, as we can see in Fig. \ref{fig:summary}g and Fig. \ref{fig:summary}h. This happens because the boundary solution $\varphi_{+}$ approaches the antikink shape as $H \to -1$ (see the Fig. \ref{fig:solutions}b). 
In fact, for $H\approx -1$, the dynamics boils down to kink-antikink scattering with reflective boundary at $x=0$. In this way, the kink-antikink collision can induce both bion creation and bounces, as can be seen in Fig. \ref{fig:summary}h and Fig. \ref{fig:summary}i, respectively. 

%%%%%%%%%%%%%%%%%%%%%%%%%%%%%%%%%%%%%%%%%%%%%%%%%%%%%%%%%%%%%%%%%%%%%%%%%%%%%%%%%%

\section{Perturbations and the boundary modes} \label{sec4}
%%%%%%%%%%%%%%%%%%%%%%%%%%%%%%%%%%%%%%%%%%%%%%%%%%%%%%%%%%%%%%%%%%%%%%%%%%%%%%%%%%
Introducing small perturbations around the static kink solution in planar wave form,
\begin{equation}
    \varphi ( x,t) =\varphi_K ( x) +\Re\{ e^{i\omega t} \psi ( x)\}
    \label{eq:perturbation}
\end{equation}
we are able to find a Schr\"odinger-like equation
\begin{equation}
    \left[ -\partial _{x}^{2} +U( x)\right] \psi ( x) =\omega ^{2} \psi ( x),
\end{equation}
where $U(x)= \left.V''(\varphi)\right|_{\varphi=\varphi_{K}}$ is the linearized potential. For the full-line analysis, $U(x)$ takes the form of a P\"oschl--Teller potential, and the eigenvalue problem reveals two discrete modes: the translational $\omega_0=0$, and the internal shape mode $\omega_1=\sqrt{3}$ . Furthermore, there is a continuous spectrum for $2 < \omega < \infty$, associated to scattering states
\begin{equation}
    \psi ( x) =e^{ikx}\left( 3\tanh^{2} x-k^{2} -1-3ik\tanh x\right),
    \label{eq:scatt}
\end{equation}
whose the dispersion relation is $\omega^2 = 4 + k^2$. 

%%%%%%%%%%%%%%%%%%%%%%%%%%%%%%%%%%%%%%%%%%%%%%%%%%%%%%%%%%%%%%%%%%%%%%%%%%%%%%%%%%
\begin{figure}[t]
    \centering
    \includegraphics[width=0.5\textwidth]{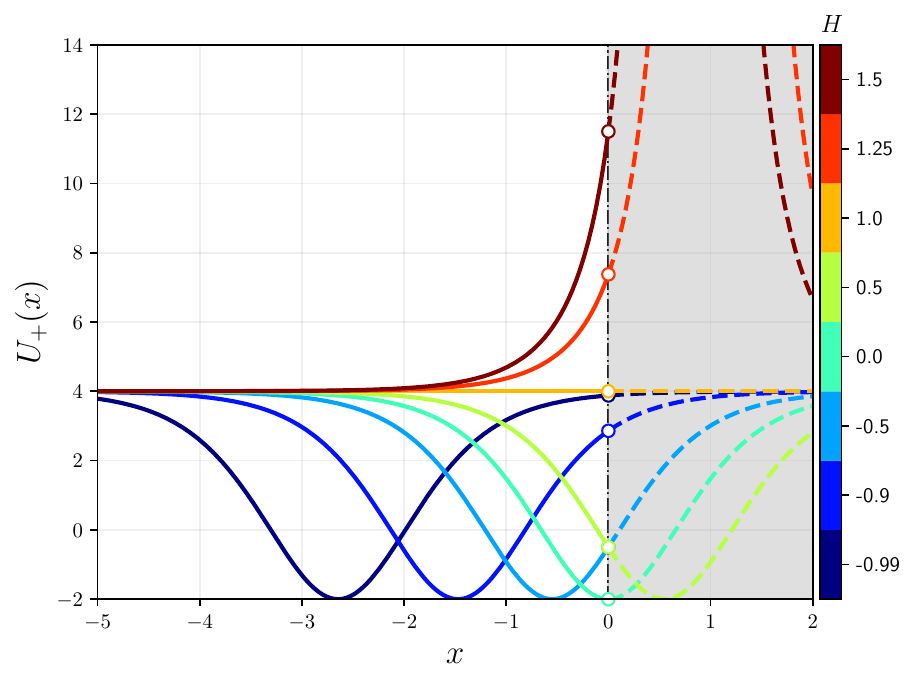}
    \caption{The Schrodinger potential for the positive boundary solution $U_{+}(x)$ and some values of $H$.}
    \label{pot_sch}
\end{figure}
%%%%%%%%%%%%%%%%%%%%%%%%%%%%%%%%%%%%%%%%%%%%%%%%%%%%%%%%%%%%%%%%%%%%%%%%%%%%%%%%%%

 By restricting the system to the half-line and imposing the Dirichlet condition in Eq. \eqref{eq:perturbation}, the scattering states for $|H|<1$ must satisfy $\psi(\chi)=0$. This results in a relationship $k_\pm=i\kappa_\pm$ for
\begin{equation}
    \kappa _{\pm } =-\frac{3}{2}H \pm \sqrt{1-\frac{3}{4} H^{2}},
    \label{eq:kappa}
\end{equation}
with the corresponding values of $\omega_\pm^2 = 4 -\kappa_\pm^2$, associated to the boundary solutions $\varphi_\pm$. The case $|H|\geq 1$ represents the region of irregular solutions in the full line, given by $\coth(x+\chi)$. Then the continuous modes in the full line are not described by the Eq. \eqref{eq:scatt}.

The Schroedinger potential is presented in the Fig. \ref{pot_sch} for $U_+(x)=\left.V''(\varphi)\right|_{\varphi=\varphi_{+}}$. For $|H|\geq 1$ the Schrodinger potential assumes a form of an exponential tail, whereas for $|H|<1$ we have a well near to the boundary.  For $U_-(x)=\left.V''(\varphi)\right|_{\varphi=\varphi_{-}}$ we have $U_-(x;H)=U_+(x;-H)$. Note that the presence of the potential well is crucial for the occurrence of bound states from the boundary for $|H|<1$ and such modes are absent for $|H|\geq1$.

%%%%%%%%%%%%%%%%%%%%%%%%%%%%%%%%%%%%%%%%%%%%%%%%%%%%%%%%%%%%%%%%%%%%%%%%%%%%%%%%%%
\begin{figure}[t]
    \centering
    \includegraphics[width=\textwidth]{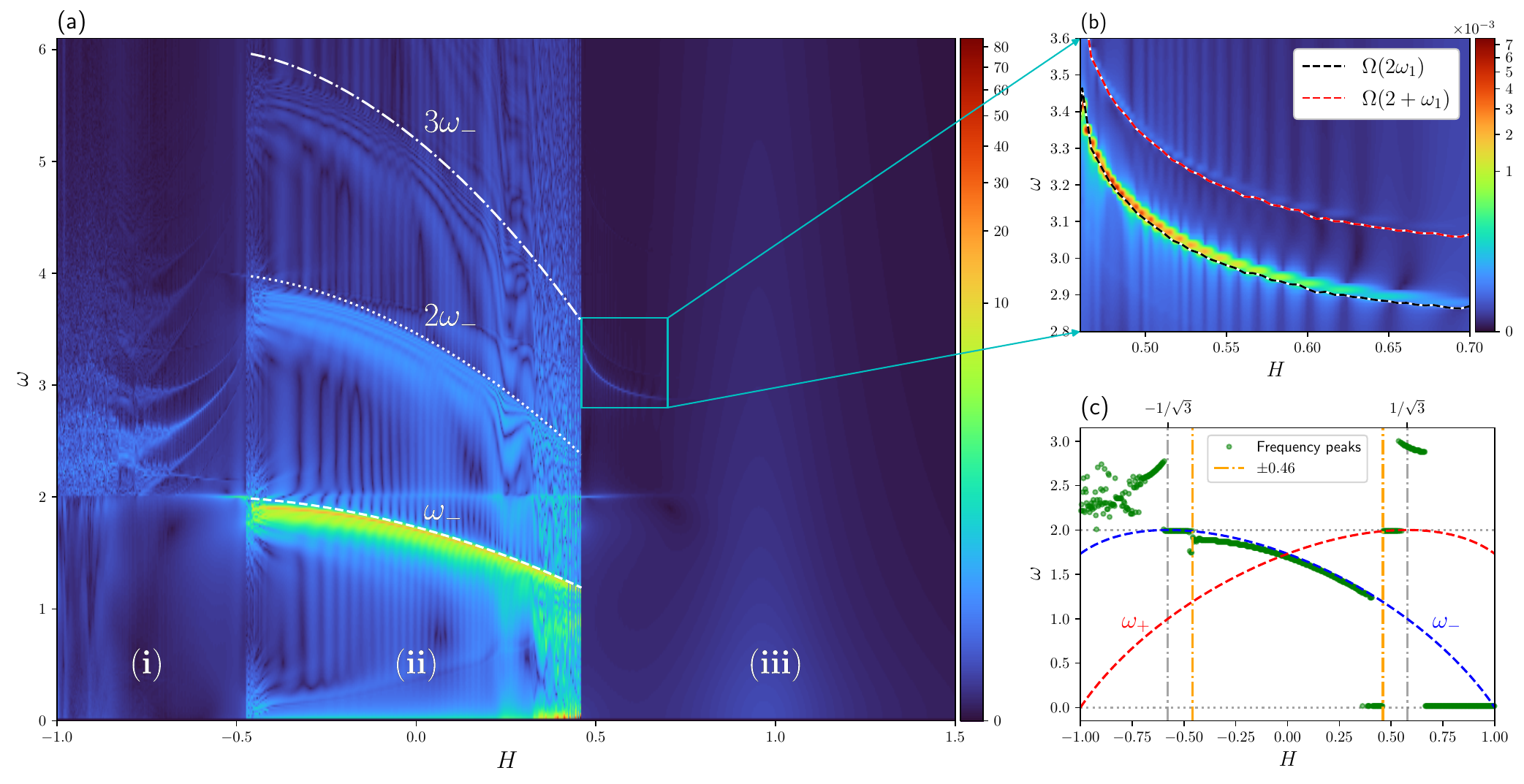}
    \caption{(a) The power spectral density of $\phi'(x,t)$ at $x=0$ for a time interval $|x_0|/v \leq t \leq 400$ and impact velocity $v=0.25$. Traced curves correspond to $\omega_-$ and its second and third harmonics. (b) A zooming-in region showing the frequencies of Doppler-shifted radiation released by the free kink after one-bounce. (c) Frequencies of the peaks with the largest spectral density of the Fig. \ref{fig:spectrum}a. }
    \label{fig:spectrum}
\end{figure}
%%%%%%%%%%%%%%%%%%%%%%%%%%%%%%%%%%%%%%%%%%%%%%%%%%%%%%%%%%%%%%%%%%%%%%%%%%%%%%%%%%

The Fig. \ref{fig:spectrum}a shows the power spectral density of $\phi'(x,t)$ at the boundary, after scattering has occurred, in the phase space $\omega$ {\it versus} $H$, for the particular case $v=0.25$. The figure presents a complex pattern with three distinct regions: (i) $-1<H\lesssim-0.46$, (ii) $-0.46\lesssim H\lesssim0.46$, (iii) $0.46\lesssim H\lesssim1.5$.
The border in $H$ between regions (ii) and (iii) corresponds to the critical velocity $v_c$, which also corresponds to the frontier between regions (II) and (III) of the Fig. \ref{fig:mosaico}a.

First of all, note the numerical evidence of the oscillating mode for the negative boundary $\omega_{-}$ (white dashed curve) in the collisions. We can see in the Fig. \ref{fig:spectrum}a, that this mode is found in the interval (ii), which corresponds to kink annihilation situation (see also the region (III) from the Fig. \ref{fig:mosaico}a), when most of the system's energy is exchanged in perturbations of the negative boundary $\varphi_{-}$. Note also in the Fig. \ref{fig:spectrum}a the presence of the second ($2\omega_{-}$) and third ($3\omega_{-}$) harmonics for part of this interval of $H$. One can see that the harmonics are degraded for higher values of $H$, where  a transition region appears for $H_c\simeq0.46$ and $\omega\simeq0$ that grows for higher $\omega$. A stochastic pattern appears in this region, highlighting the emission spectrum of the kink-boundary bound state.  Another aspect of the region (ii) appears for $H \gtrsim -0.46$, where a small discrepancy  appears between the resonant peaks and the mathematical expression of $\omega_-$. This is more evident for the second and third harmonics.

For values of $H$ above the critical value, $H > H_c$, we can observe in the Fig. \ref{fig:spectrum}a the region $(iii)$ with a much cleaner spectrum, with three frequency peaks: one in the threshold frequency $\omega=2$ and the other two above. In the Fig. \ref{fig:spectrum}b we show a zoom in of the Fig. \ref{fig:spectrum}a. Note that this region corresponds roughly to the region (III) from the Fig. \ref{fig:mosaico}a, where there in inelastic scattering (one-bounce). This means scatterings where there is a scattered kink, as illustrated, for instance, in the Fig. \ref{fig:summary}b. The vibrational mode $\omega_1$ decays before detected at the boundary. Following the procedure developed in the Ref. \cite{dor}, we can see that in this case the two frequency peaks above the threshold frequency correspond to Doppler-shifted frequencies $\Omega(\omega) = \gamma_f(\omega + k(\omega)v_f)$ of the radiation released from the scattered kink measured at the boundary. One such mode is the second harmonic $(2\omega_1)$, detected in the boundary at frequency $\Omega(2\omega_1)$. The other detected frequency corresponds to the combination of the threshold frequency (the minimum frequency related to regions close to the boundary) and the internal mode, $\Omega(2+\omega_1)$. See these two frequencies in the Fig. \ref{fig:spectrum}b. Note also that, as we increase $H$, these two peaks disappear in such way that we have only translational modes for $H \gtrsim 1$, characterizing the elastic scattering.

In the Fig. \ref{fig:spectrum}c we show the peaks with largest spectral density of the Fig. \ref{fig:spectrum}a. Note from the figure the presence of the mode $\omega=0$. The presence of this mode is related to the occurrence of a free moving kink for the following situations: multi-bounce kink-boundary scattering (Fig. \ref{fig:summary}d), when $0.4\lesssim H \lesssim 0.46$; one-bounce, when $1/\sqrt{3} \lesssim H \lesssim 1$. This mode also occurs for $H\sim-1$, where the boundary is like an antikink (Fig. \ref{fig:summary}i), which is more visible for high values of $v$. Note the presence of $\omega_-$ for $-1/\sqrt{3}\lesssim H\lesssim 0.3$, corresponding to excitation near the boundary (Fig. \ref{fig:summary}e), and the production of bions for $-1<H<-1/\sqrt{3}$ (Fig. \ref{fig:summary}h). For $0.46 < H \lesssim 0.6$, the more significant points correspond to Doppler and threshold frequency $\omega=2$. Note that the spectrum for $H \lesssim -0.5$ does not show excitation of the boundary mode $\omega_{-}$. As we can see in the Figs. \ref{fig:spectrum}a and \ref{fig:spectrum}c, this region revels only perturbations above the threshold frequency $\omega > 2$, which characterize the oscillon and bion emission: oscillon signature in $-0.75 \lesssim H \lesssim -0.5$; and the stochastic bion spectrum at $H \lesssim -0.75$.

%%%%%%%%%%%%%%%%%%%%%%%%%%%%%%%%%%%%%%%%%%%%%%%%%%%%%%%%%%%%%%%%%%%%%%%%%%%%%%%%%%
\begin{figure}[t]
    \centering
    \includegraphics[width=0.9\textwidth]{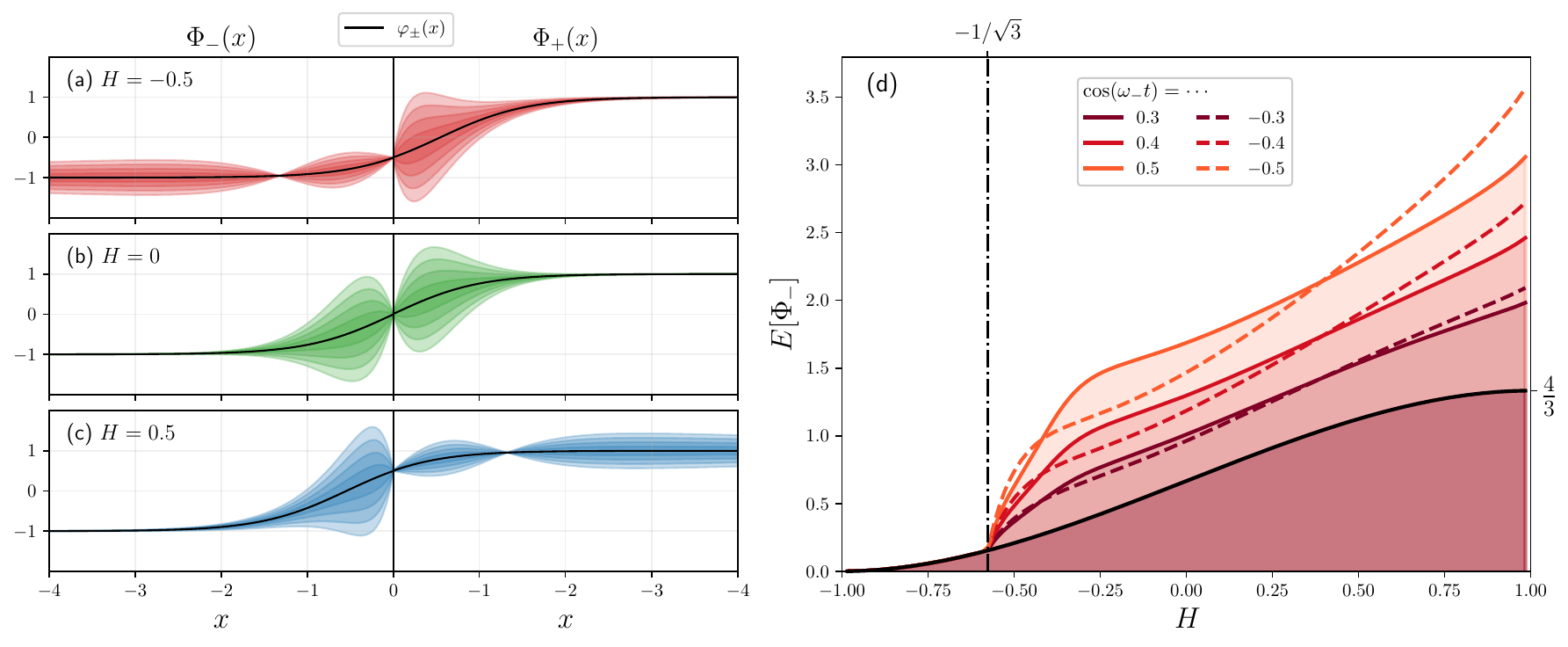}
    \caption{Vibrational shape mode $\Phi_\pm(x)$ with (a) $H=-0.5$, (b) $H=0$ and (c) $H=0.5$ for some values of $\cos(\omega_-t)$. (d ) Static energy of the perturbed solution $\Phi_-(x)$ as function of $H$ for some values of $\cos(\omega_-t)$.}
    \label{perturb}
\end{figure}
%%%%%%%%%%%%%%%%%%%%%%%%%%%%%%%%%%%%%%%%%%%%%%%%%%%%%%%%%%%%%%%%%%%%%%%%%%%%%%%%%%

Let us now consider the occurrence of shape modes $\tilde\varphi_\pm(x)$ related to each of the boundary solutions $\varphi_\pm(x)$. Then the boundary solutions would have the following form
\be 
\Phi_\pm(x,t)=\varphi_\pm(x) +   \cos(\omega_\pm t) \tilde\varphi_\pm(x) . 
\ee
Let us consider the shape mode as $\tilde{\varphi }_{\pm}( x) = \mathcal{N}_\pm\psi _\pm( x)$, where
\begin{equation}
    \psi _\pm( x) =e^{\kappa _\pm ( \chi \pm x)}\left( 3\tanh^{2}( \chi \pm x) +\kappa _\pm^{2} -1+3\kappa _\pm\tanh( \chi \pm x)\right)
    % \label{eq:negative boundary perturbation *}
\end{equation}
satisty the Dirichlet condition, $\psi_\pm(0)=0$ and 
\begin{equation}
    \mathcal{N}_\pm(H)=\frac{1}{\sqrt{\int_{-\infty}^0 |\psi_\pm(x;H)|^2\,dx}}
    \label{eq:norm}
\end{equation}
are the respective normalization constants. Note that the asymptotic behavior for $\psi _\pm( x)$ is governed by the exponential term. Therefore, for $\pm\kappa_\pm \geq 0$ the normalization assumes a positive non-null value and then we have two shape modes, one for each solution $\varphi_{\pm}$ which oscillates near the boundary.
In the Figs. \ref{perturb}a-c we present the shape modes for some values of $H$. Note from the figure that the shape mode is localized around the boundary, and, for symmetry, $\tilde{\varphi}_+(x)|_H=\tilde{\varphi}_-(x)|_{-H}$. Note that for $\pm\kappa_\pm < 0$, the integral in Eq. (\ref{eq:norm}) diverges, suppressing the respective boundary shape mode. We can see this effect in the Fig. \ref{perturb}d. Consequently, the boundary oscillation frequency $\omega_{-}$ can only be excited for $H>-1/\sqrt{3}$, as is illustrated in the Fig. \ref{fig:spectrum}c. In the same way, $\omega_{+}$ can only appear for $H<1/\sqrt{3}$ (see the Fig. \ref{fig:spectrum}c and next Figs. \ref{fft}c-d).

The static energy  of the perturbed negative boundary is given by 
\be
E[ \Phi_- ] =\int\nolimits _{-\infty }^{0} dx\left(\tfrac{1}{2} \Phi ^{\prime 2} +V( \Phi )\right).
\ee
In the Fig. \ref{perturb}d we present $E[{\Phi}_{-}]$ as a function of $H$ for several fixed values of $\cos(\omega_-t)$. Note that when the static energy of $\Phi_-(x)$ converge to the energy of the non-perturbed solution $\varphi_-(x)$ as $H \to -1/\sqrt{3}$ for any value of $\cos(\omega_-t)$.

%%%%%%%%%%%%%%%%%%%%%%%%%%%%%%%%%%%%%%%%%%
\begin{figure}[t]
    \centering
    \includegraphics[width=\textwidth]{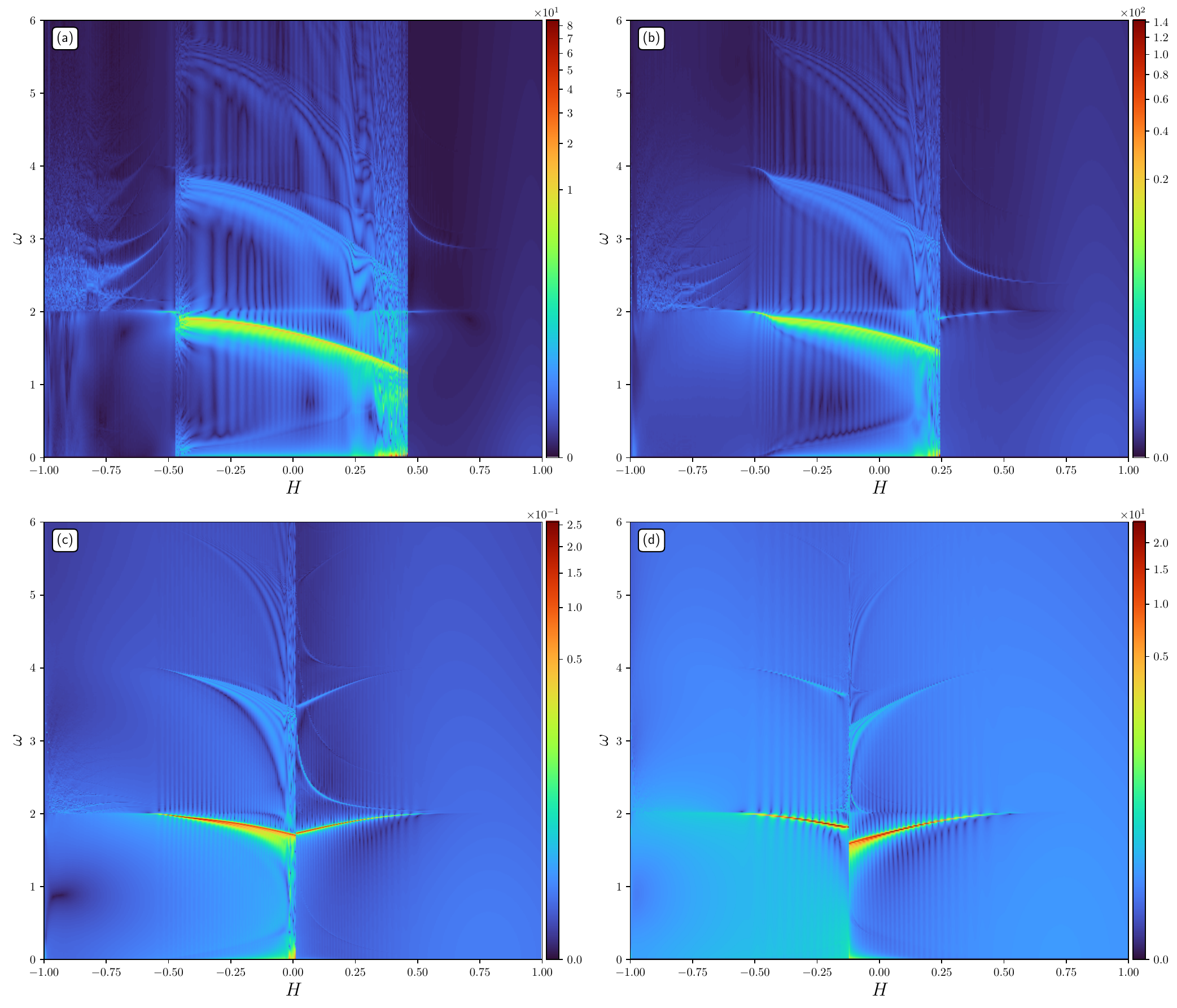}
    \caption{ The power spectral density of $\phi'(x,t)$ at $x=0$ for a time interval $|x_0|/v \leq t \leq 400$ and impact velocities a) $v=0.25$, b) $v=0.5$, c) $v=0.75$, d) $v=0.9$. }
    % \label{fig:amplitude_modulation}
    \label{fft}
\end{figure}
%%%%%%%%%%%%%%%%%%%%%%%%%%%%%%%%%%%%

A similar analysis shows that the frequency $\varphi_{+}$ only can be perturbed for $H < 1/\sqrt{3}$.  Observing the Fig. \ref{fig:spectrum}c, we can clearly see that the boundary modes $\omega_\pm$ only are excited for $-1/\sqrt{3} < H < 1/\sqrt{3}$. These aspects can be seen in the Fig. \ref{fft}, where is shown the spectrum for various values of $v$. In the Fig. \ref{fft}a we see the presence of the threshold $\omega=2$ that cannot be distinguished from $\omega_+$ (the same appears in Fig. \ref{fig:spectrum}c). With the increasing of $v$, the influence of $\omega_+$ is more evident, with strong peaks for $v=0.9$. The contribution of $\omega_-$ and $\omega_+$ is similar for $H=0.75$ and $H=0.9$. Note also that the transition region, where occurs the production of two-bounce kink-boundary is reduced with the increasing of $v$. This can be compared with the Fig. \ref{fig:mosaico}, where the granularity of the regions close to $v_c$ decrease with the increasing of $v$, until a point where the two-bounce disappears.

The Figs. \ref{fft}a-d show the power spectral density of $\phi'(x,t)$ at the boundary after scattering for four different impact velocities. Note that for a small velocity (Figs. \ref{fft}a for $v=0.25$) the mode $\omega_-$ is favoured. On the contray, for a large velocities ($v=0.75$ and $v=0.9$ from Figs. \ref{fft}c-d) shows that both $\omega_+$ and $\omega_-$ have high spectral density.    
%%%%%%%%%%%%%%%%%%%%%%%%%%%%%%%%%%%%%%%%%%%%%%%%%%%%%%%%%%%%%%%%%%%%%%%%%%%%%%%%%%

\section{Resonance scatterings} \label{sec5}
%%%%%%%%%%%%%%%%%%%%%%%%%%%%%%%%%%%%%%%%%%%%%%%%%%%%%%%%%%%%%%%%%%%%%%%%%%%%%%%%%%
In this section, we present three interesting resonant scatterings. 
Two of them, related to soft oscillations and oscillons, we think have not been reported before in the $\phi^4$ model.

%%%%%%%%%%%%%%%%%%%%%%%%%%%%%%%%%%%%%%%%%%%%%%%%%%%%%%%%%%%%%%%%%%%%%%%%%%%%%%%%%%

\subsection{The kink-boundary resonance mechanism}

%%%%%%%%%%%%%%%%%%%%%%%%%%%%%%%%%%%%%%%%%%%%%%%%%%%%%%%%%%%%%%%%%%%%%%%%%%%%%%%%%%

%%%%%%%%%%%%%%%%%%%%%%%%%%%%%%%%%%%%%%%%%%%%%%%%%%%%%%%%%%%%%%%%%%%%%%%%%%%%%%%%%%
\begin{figure}
    \centering
    \includegraphics[width=\linewidth]{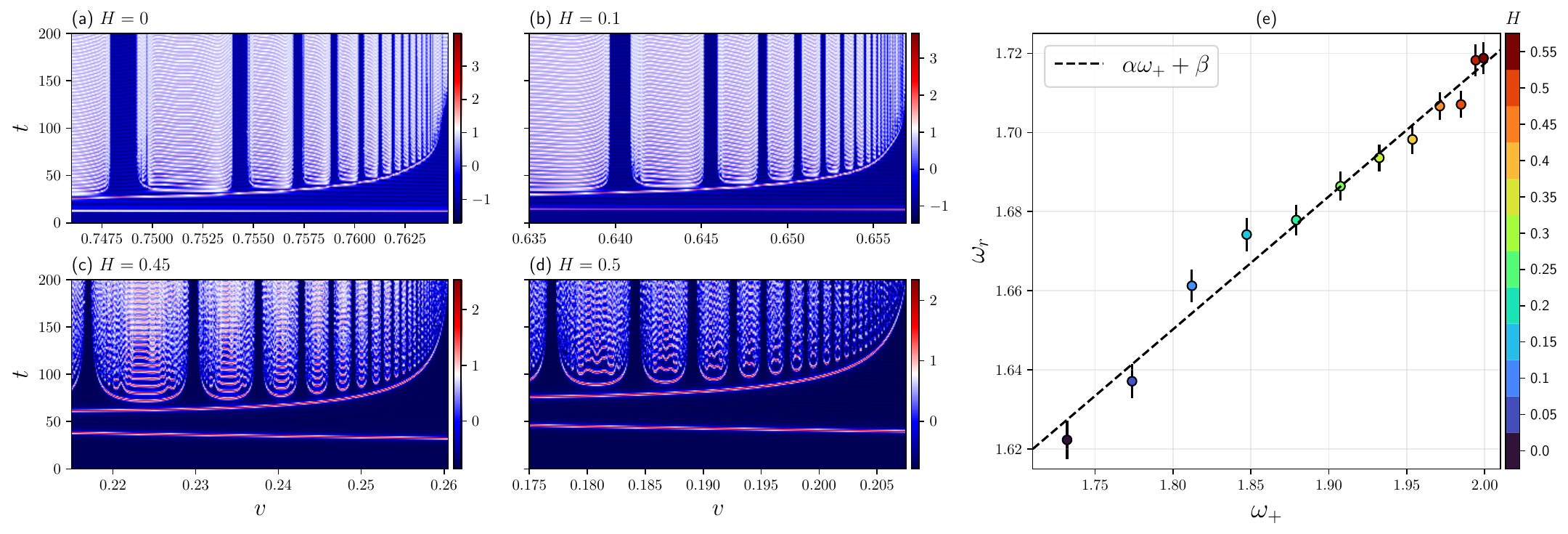}
    \caption{(a)-(d) Examples of resonance windows in the kink-boundary interaction; (e) the linear tendency between the resonance frequency $\omega_r$ and the positive boundary frequency $\omega_{+}$ and the linear fitting (black dashed line).}
    \label{fig:resonance}
\end{figure}
%%%%%%%%%%%%%%%%%%%%%%%%%%%%%%%%%%%%%%%%%%%%%%%%%%%%%%%%%%%%%%%%%%%%%%%%%%%%%%%%%%

 The figures \ref{fig:resonance}a-d show the presence of resonance windows for kink-boundary scattering for different values of $H$.  The scattering results show that, for $v \approx v_c$, we can have situations analogous to the two-bounce phenomenon between the kink and antikink in the full line. An example of this type of two-bounce scattering can be seen in the Fig. \ref{fig:summary}d. Note that the critical velocity is reduced with the increasing of $H$, favoring the one-bounce scattering. In the full-line theory, the multi-bounce windows phenomenon is triggered by the resonance mechanism \cite{k1}
\begin{equation}
    \omega_r \Delta t = 2\pi m + \delta,
\end{equation}
in which the resonance frequency is the kink shape mode $\omega_r=\omega_1$ and $m$ is the field oscillation before the pair kink-antikink escape. In the half-line, the situation is more complex, with different resonance mechanisms in action depending on the values of the parameters. The numerical simulations reveal a dependency on the resonance frequency $\omega_r$ and the Dirichlet boundary mode. As one can see in the Fig. \ref{fig:resonance}e, for $0 \leq H < 1$, we have a linear relation between the resonance frequency and the frequency $\omega_+$ of the boundary, of the form 
\begin{equation}
    \omega_r \approx \alpha \omega_{+} + \beta\,,
\end{equation}
where $\alpha = 0.33 \pm 0.01$ and $\beta = 1.04 \pm 0.03$.

%%%%%%%%%%%%%%%%%%%%%%%%%%%%%%%%%%%%%%%%%%%%%%%%%%%%%%%%%%%%%%%%%%%%%%%%%%%%%%%%%%

\subsection{Two-bounce of soft oscillating pulses}

%%%%%%%%%%%%%%%%%%%%%%%%%%%%%%%%%%%%%%%%%%%%%%%%%%%%%%%%%%%%%%%%%%%%%%%%%%%%%%%%%%
\begin{figure}[t]
    \centering
    \includegraphics[width=\textwidth]{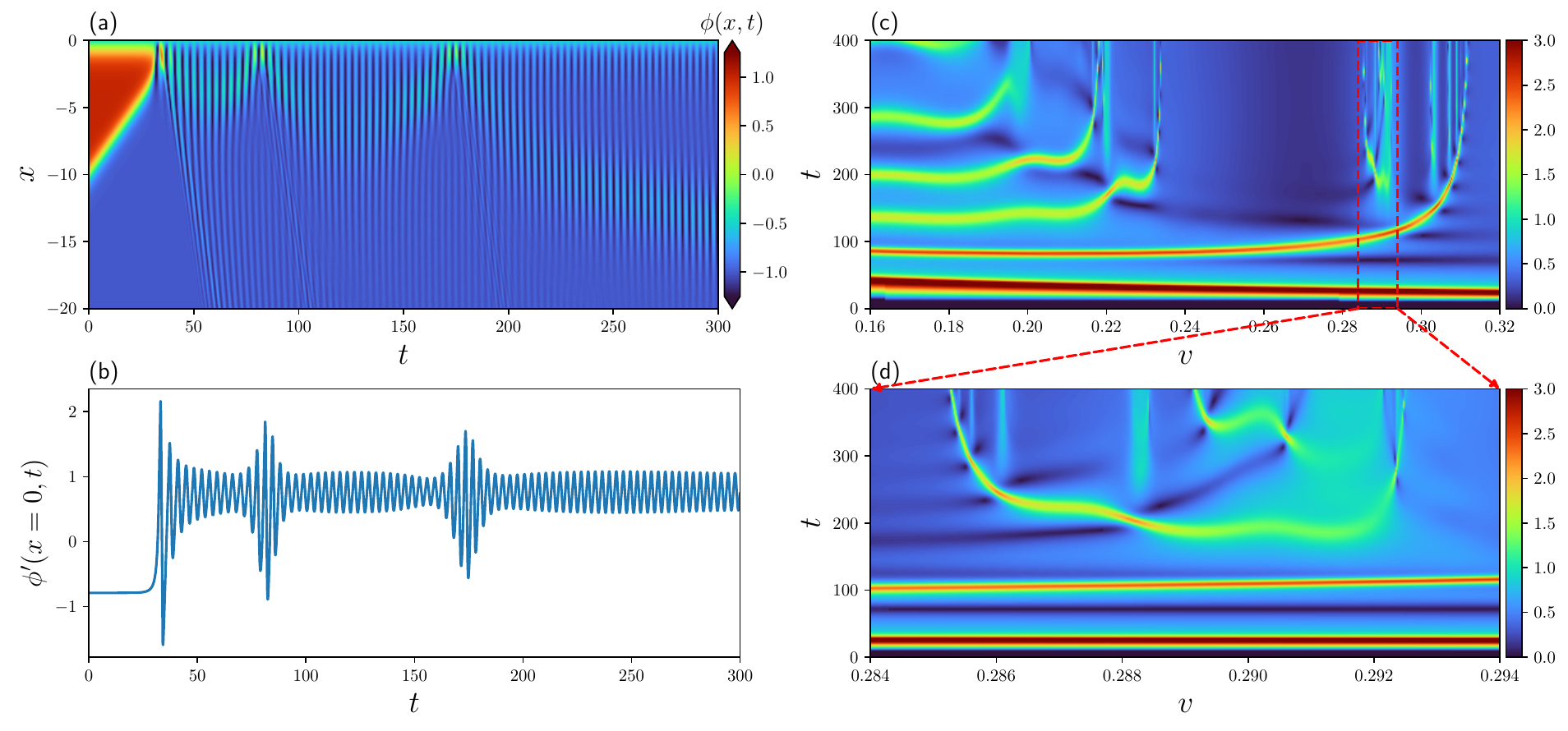}
    \caption{(a) Two-bounce interaction between the released oscillation pulses and the boundary for initial parameters $(v, H)=(0.221, -0.46)$. (b) Signal $\phi'(x,t)$ measured at $x=0$. (c) Time evolution of the amplitude modulation for some initial velocities and $H=-0.46$, and (d) a zoom in of the Fig. \ref{fig:am}c. The video depicting the scattering of figure (a) are included in the \hyperref[sec7]{Supplementary Material} as video4.mp4}.
    \label{fig:am}
\end{figure}
%%%%%%%%%%%%%%%%%%%%%%%%%%%%%%%%%%%%%%%%%%%%%%%%%%%%%%%%%%%%%%%%%%%%%%%%%%%%%%%%%%

In this section, we will present a qualitative description of low-energy oscillating pulse dynamics. For low incoming velocities ($v \lesssim 0.5$), the collisions can show the release of a oscillating lump, as is illustrated in the Fig. \ref{fig:summary}g for $H = -0.5$. Also, it can return and collide again once (Fig. \ref{fig:summary}f), twice (Fig. \ref{fig:am}a) or multiple times, revealing an analogue of multi-bounce structure in the usual kink-antikink scattering. 
This type of dynamics was reported in Ref. \cite{were_osc}, where the initial configuration generates an excited oscillation, that subsequently decays into a oscillon pair, which collide a few times and then escape from the mutual attraction.

The Fig. \ref{fig:am}a shows the modulation amplitude for $v=0.221$ and $H=-0.46$. In this case, an oscillation pulse emerges and scatters out following the kink initial impact with the boundary. However, it is unable to evade the attraction due to lack energy, resulting in two additional collisions with the boundary. The behavior of the two-bounce interaction between the oscillon and the boundary can also be illustrated in the Fig. \ref{fig:am}b, which represents the signal $\phi'(x,t)$  at $x=0$ as a function of time. Furthermore, the Fig. \ref{fig:am}c shows the evolution of the amplitude modulation for various values of $v$ with $H=-0.46$. This figure has several horizontal lines that illustrate the collisions with the boundary. A red line represents the interaction for the kink, whereas a green line represents a collision of the oscillating pulses with the boundary.  The blue region indicates that the oscillation pulses propagate off the boundary. From the figure we see that we have intervals in velocity characterizing n-bounce windows, with $1\leq n\leq 5$ for the generated oscillating pulses. Note that lower values of $v$ favor a large number of bounces. However, the presence of an even larger number of bounces is suppressed due to the emission of scalar radiation and dispersion of the oscillating pulses. For larger velocities, the bounce windows are thinner, which needs a refinement in $v$ to be observed. This is done in the Fig. \ref{fig:am}d, which is a zoom in of part the Fig. \ref{fig:am}c. This figure allows us to view one-, two-, and three-bounce windows.

%%%%%%%%%%%%%%%%%%%%%%%%%%%%%%%%%%%%%%%%%%%%%%%%%%%%%%%%%%%%%%%%%%%%%%%%%%%%%%%%%%
\subsection{Creation of kinks near the boundary}
%%%%%%%%%%%%%%%%%%%%%%%%%%%%%%%%%%%%%%%%%%%%%%%%%%%%%%%%%%%%%%%%%%%%%%%%%%%%%%%%%%

%%%%%%%%%%%%%%%%%%%%%%%%%%%%%%%%%%%%%%%%%%%%%%%%%%%%%%%%%%%%%%%%%%%%%%%%%%%%%%%%%%
\begin{figure}[t]
    \centering
    \includegraphics[width=\textwidth]{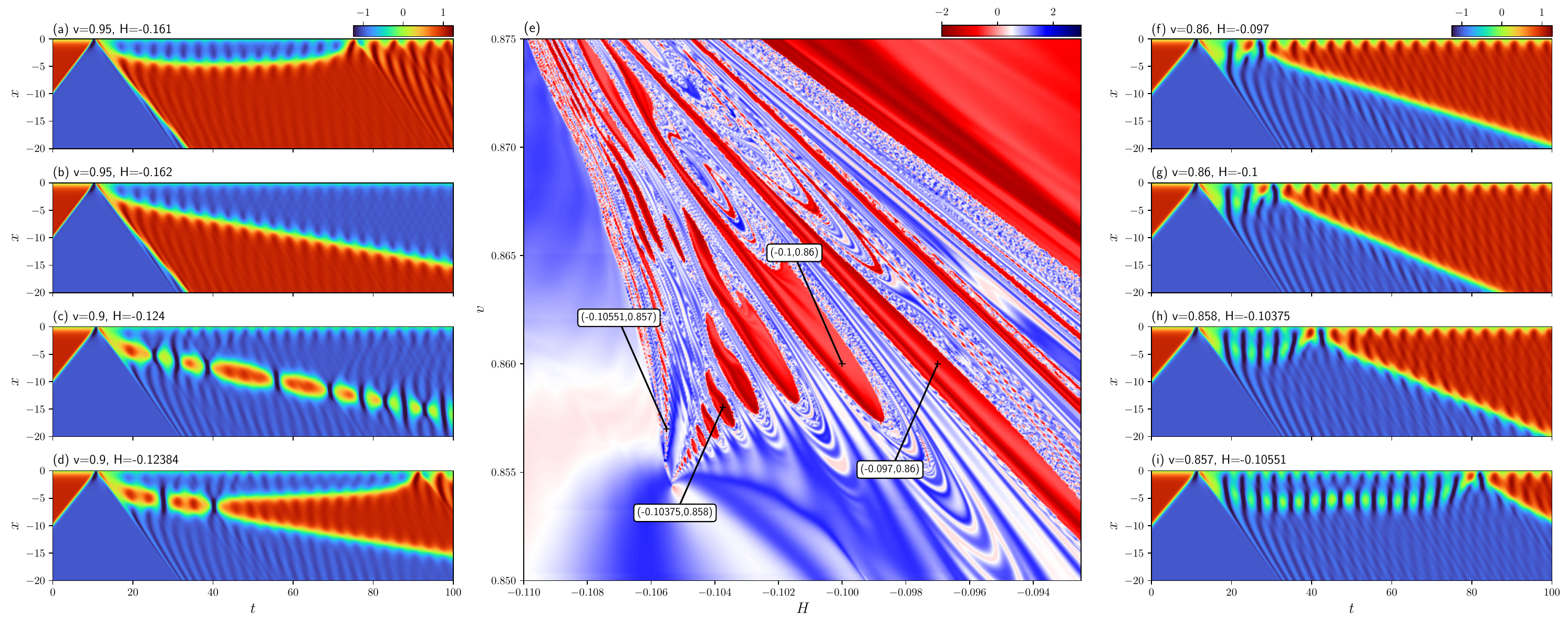}
    \caption{(a)-(d) Creation of an antikink and some aspects of its dynamics. (e) A zooming-in in the Fig \ref{fig:mosaico}a showing windows where the kink is generated by high oscillations near the boundary, with tags indicating the parameter location of the situations (f)-(i) in which the kink creation is induced by oscillating pulses. The videos depicting the scattering of figures (d) and (i) are included in the \hyperref[sec7]{Supplementary Material} as video5.mp4 and video6.mp4, respectively.}
    \label{fig:kink creation}
\end{figure}
%%%%%%%%%%%%%%%%%%%%%%%%%%%%%%%%%%%%%%%%%%%%%%%%%%%%%%%%%%%%%%%%%%%%%%%%%%%%%%%%%%

For high impact velocities, $v \gtrsim 0.8$, the excitation of the field near the boundary can generate antikinks or kinks. Those situations occur for $-0.2 \lesssim H \lesssim -0.1$. In the context of one-bounce kink-boundary scattering, high velocity kinks can produce boundary perturbations whose energy is sufficient to create an antikink high after the collision. The released antikink can: return to boundary (Fig. \ref{fig:kink creation}a), escape with a constant velocity (Fig. \ref{fig:kink creation}b) or interact with the scattered kink (Figs. \ref{fig:kink creation}c-d). 

Reducing the velocity, we found the emission of a high energy oscillon, which can be attracted back to the boundary, interfering constructively and enabling the resurgence of a free kink. This can be seen in the Figs. \ref{fig:kink creation}f-i. This is analogous to what was presented in the Ref. \cite{simashora}, where an antikink-kink pair is created after the decay of an oscillon.

The Fig. \ref{fig:kink creation}e is a zoom in of the  Fig. \ref{fig:mosaico}a. In the regions in red there is the formation of kink. The figure shows that this phenomenon appears in a complex fractal structure.
Also, for fixed $H$, the regions acquire a structure of windows in velocity. Each window can be characterized by the number $\ell$ of oscillon pulsations before the interference and emission of a kink. In this way, the emitted oscillon is bound to the boundary region since it cannot escape from his attraction. For example, in the Figs. \ref{fig:kink creation}f-i we show situations in which the oscillon has pulsations with $\ell=1,2,5$ and $15$, respectively. For the Figs. \ref{fig:kink creation}f-h, both oscillon and near boundary region have the same frequency. For the Fig. \ref{fig:kink creation}i, on the contrary, we have $\ell=15$ for the oscillon, whereas the boundary oscillates sixteen times. When this occurs, it can be traced in blue regions of the oscillon, that can present one or more bifurcations in the direction to the boundary. Also, the Figs. \ref{fig:kink creation}f-h show that the oscillon escapes and returns to the boundary, having a profile of a one-bounce collision with the boundary. On the other hand, the emitted oscillon from the Fig. \ref{fig:kink creation}i have an envelope presenting a smooth oscillation in the middle region, showing the tendency of a two-bounce collision with the boundary.    

The configurations of the Figs. \ref{fig:kink creation}f-i can be localized in the phase space diagram of the Fig. \ref{fig:kink creation}e. In that figure one can see that the red regions characterize the number of oscillations. The higher the number of oscillations, the smaller the region in which they occur. The red regions are reduced and deformed for even larger values of $\ell$. The limit $\ell\to\infty$ would characterize oscillatory pulses near the boundary that would not decay. 
However, such a bound stable state is not expected to occur because of the growing deformation in the small spots imposed by dissipation and numerical precision. In this case, it is expected that the decay time for the bound oscillon-boundary state to be of the order of the decay time of the oscillon. Lastly, we can say that it is the mutual influence between the number of bifurcations from the boundary oscillations toward the oscillon and the number of oscillations that gives rise to the fractal pattern described in the Fig. \ref{fig:kink creation}e.

%%%%%%%%%%%%%%%%%%%%%%%%%%%%%%%%%%%%%%%%%%%%%%%%%%%%%%%%%%%%%%%%%%%%%%%%%%%%%%%%%%
\section{Conclusions} \label{sec6}
%%%%%%%%%%%%%%%%%%%%%%%%%%%%%%%%%%%%%%%%%%%%%%%%%%%%%%%%%%%%%%%%%%%%%%%%%%%%%%%%%%

In this work, we have considered the $\phi^4$ kink in the half-line with a Dirichlet boundary condition. Adjusting the spatial displacement, we showed that some irregular solutions in the full line satisfy the boundary condition, which enriches the possibilities of kink-boundary scattering, depending on the parameters $(H,v)$.
Some interesting possibilities reported are elastic scattering; one-bounce scattering;  bound-state between the kink and boundary; two-bounce scattering; total annihilation of the
kink, in which case most of the energy decays in boundary perturbations; releasing of an oscillon that suffers itself a two-bounce scattering with the boundary; emission of an oscillon; creation of an antikink, which interacts with the incoming kink, resulting in a moving bion; three-bounce collision with the emission of a kink-antikink pair. 

The main scenarios of scattering are distributed is a map of the first spatial derivative of the field at the boundary interns of the parameters $(H,v)$. We showed an almost uniform distribution for elastic scattering for $1\lesssim H\lesssim1.5$ and oscillon creation for $-1\lesssim H\lesssim -0.5$. A more complicated pattern is observed for inelastic scattering (one-bounce) and boundary-induced annihilation, with evidence of a linear frontier between these two regions. 

 Linear perturbation of the kink-like boundary solution results in two shape modes. A good way to see the distribution of the modes in the scattering process is through the power spectral density of $\phi'(x, t)$ at the boundary after scattering has occurred, in the phase space $\omega$ versus $H$. This was done for a particular value of $v$. We explain some spectral lines, such as the boundary mode and those related to Doppler-shifted radiation released by the free kink after suffering a one-bounce collision. The values of $H$ that favor the production of oscillons are analyzed considering the energy of the perturbed boundary, showing that there is a minimum that favors the annihilation of the incident kink.

We identified three processes of n-bounce scatterings: i) the kink-boundary resonance mechanism, analogous to the two-bounce phenomenon of the kink-antikink scattering in the full line. Here the resonance frequency is a linear function of the oscillation mode of the boundary.
ii) the low-energy and highly dispersive oscillating pulses emitted for kinks with low incoming velocity. 
iii) the creation of antikinks or kinks for high impact velocities and a short interval of $H$. Despite representing a small region in phase space, the structure of resonance observed for the intermediate oscillating pulses is very rich, presenting a self-similar structure. 

In summary, the simple $\phi^4$ model has shown to be an extremely useful model to describe the influence of nonlinearity and nonintegrability in the dynamics, even in the half-line with the simplest Dirichlet boundary.

%%%%%%%%%%%%%%%%%%%%%%%%%%%%%%%%%%%%%%%%%%%%%%%%%%%%%%%%%%%%%%%%%%%%%%
\section{Supplementary Material} \label{sec7}
%%%%%%%%%%%%%%%%%%%%%%%%%%%%%%%%%%%%%%%%%%%%%%%%%%%%%%%%%%%%%%%%%%%%%%

\begin{itemize}
\item \href{link here}{\texttt{video1.mp4}} - Collision for $(v,H)=(0.25, 2)$, revealing a elastic scattering.

\item \href{link here}{\texttt{video2.mp4}} - Example of two-bounce in the kink-boundary collision, for $(v,H)=(0.25, 0.44)$.

\item \href{link here}{\texttt{video3.mp4}} - Case for $(v,H)=(0.25, 0)$ in which the kink is annihilated, and most of its energy is exchanged into perturbations near boundary.

\item \href{link here}{\texttt{video4.mp4}} - Example of two-bounce in the oscillon-boundary collision, for $(v,H)=(0.221,-0.46)$.

\item \href{link here}{\texttt{video5.mp4}} - Collision for $(v,H)=(0.9,-0.12384)$, in which the kink-boundary collision generates a kink-antikink pair, that interacts forming a two-bounce situation.

\item \href{link here}{\texttt{video6.mp4}} - Collision for $(v,H)=(0.857,-0.10551)$, whose result is a static oscillating pulse, that return after fifteen times and induces the creation of a kink.
\end{itemize}

%%%%%%%%%%%%%%%%%%%%%%%%%%%%%%%%%%%%%%%%%%%%%%%%%%%%%%%%%%%%%%%%%%%%%%
\section*{Acknowledgements}
%%%%%%%%%%%%%%%%%%%%%%%%%%%%%%%%%%%%%%%%%%%%%%%%%%%%%%%%%%%%%%%%%%%%%%

 The authors thank the anonymous referee for suggestions, in particular concerning the description of the presence of shape modes. The authors thank K.Z. Nobrega for discussions. F.C. Simas thanks FAPEMA - Funda\c c\~ao de Amparo \`a Pesquisa e ao Desenvolvimento do Maranh\~ao through Grants COOPI-07838/17. 
A.R. Gomes thanks UFMA through Grants PVCET3471-2023, FAPEMA - Funda\c c\~ao de Amparo \`a Pesquisa e ao Desenvolvimento do Maranh\~ao through Grants Universal 01441/18, COOPI-07838/17 and CNPq (brazilian agency) through Grants 313014. This work was partially supported by the Coordena\c c\~ao de Aperfei\c coamento de Pessoal de
N\'ivel Superior - Brasil (CAPES) - Finance Code 001.

%%%%%%%%%%%%%%%%%%%%%%%%%%%%%%%%%%%%%%%%%%%%%%%%%%%%%%%%%%%%%%%%%%%%%%

%%%%%%%%%%%%%%%%%%%%%%%%%%%%%%%%%%%%%%%%%%%%%%%%%%%%%%%%%%%%%%%%%%%%%%%%%%%%%%%%%%

\end{document}